\documentclass[sigconf]{acmart}
\usepackage{graphicx}
\usepackage{caption}
\usepackage{subcaption}

\newcommand{\mat}{\mathbf}

\graphicspath{{./figures/}}
\usepackage{multirow}
\usepackage{makecell}
\usepackage{enumitem}
\newcommand{\indep}{\rotatebox[origin=c]{90}{$\models$}}
\usepackage{pifont}
\newcommand{\cmark}{\ding{51}}%
\newcommand{\xmark}{\ding{55}}%
\usepackage[colorinlistoftodos,prependcaption,textsize=tiny]{todonotes}
\usepackage{color}

\theoremstyle{assumption}
\newtheorem{assumption}{Assumption}
\theoremstyle{definition}
\newtheorem{definition}{Definition}

\AtBeginDocument{%
  \providecommand\BibTeX{{%
    \normalfont B\kern-0.5em{\scshape i\kern-0.25em b}\kern-0.8em\TeX}}}

\copyrightyear{2021}
\acmYear{2021}
\setcopyright{acmcopyright}\acmConference[WSDM '21]{Proceedings of the Fourteenth ACM International Conference on Web Search and Data Mining}{March 8--12, 2021}{Virtual Event, Israel}
\acmBooktitle{Proceedings of the Fourteenth ACM International Conference on Web Search and Data Mining (WSDM '21), March 8--12, 2021, Virtual Event, Israel}
\acmPrice{15.00}
\acmDOI{10.1145/3437963.3441719}
\acmISBN{978-1-4503-8297-7/21/03}

\begin{document}
\fancyhead{}
\title{Long-Term Effect Estimation with Surrogate Representation}
\author{Lu Cheng, Ruocheng Guo, Huan Liu}
\affiliation{%
  \institution{School of Computer Science and Engineering, Arizona State University}
}
\email{{lcheng35,rguo12,huanliu}@asu.edu}
\renewcommand{\shortauthors}{Cheng, et al.}

\begin{abstract}
There are many scenarios where short- and long-term causal effects of an intervention are different. For example, low-quality ads may increase short-term ad clicks but decrease the long-term revenue via reduced clicks. This work, therefore, studies the problem of long-term effect where the outcome of primary interest, or \textit{primary outcome}, takes months or even years to accumulate. The observational study of long-term effect presents unique challenges. First, the confounding bias causes large estimation error and variance, which can further accumulate towards the prediction of primary outcomes. Second, short-term outcomes are often directly used as the proxy of the primary outcome, i.e., the \textit{surrogate}. Nevertheless, this method entails the strong surrogacy assumption that is often impractical. To tackle these challenges, we propose to build connections between long-term causal inference and sequential models in machine learning. This enables us to learn \textit{surrogate representations} that account for the \textit{temporal unconfoundedness} and circumvent the stringent surrogacy assumption by conditioning on the inferred time-varying confounders. Experimental results show that the proposed framework outperforms the state-of-the-art.
\end{abstract}



\keywords{Long-Term Effect, Surrogates, Sequential Models, Representation Learning}

\maketitle
\section{Introduction}
We often concern about the treatment effect measured in a short time frame, i.e., \textit{short-term effect}, as it is stable and generalizable to the \textit{long-term treatment effect} \cite{kohavi2020trustworthy}. However, short- and long-term effects can differ in some scenarios.
For instance, search engines measured by inappropriate performance metrics may increase search query shares in a short-term but not in the long-term as users switch to better search engines; low-quality ads (e.g., ads with clickbait) may increase clicks and revenue in the short-term but decreases revenue via reduced ad clicks, and even searches in the long-term \cite{hohnhold2015focusing}. Because long-term outcomes, i.e., \textit{primary outcomes}, may be observed well beyond the time frame required to make policy decisions \cite{athey2019surrogate}, measuring long-term effects is challenging, especially in an online world where products and services are developed quickly and iteratively in an agile fashion \cite{kohavi2020trustworthy}.  
\begin{figure}
\center
  \includegraphics[width=.85\columnwidth]{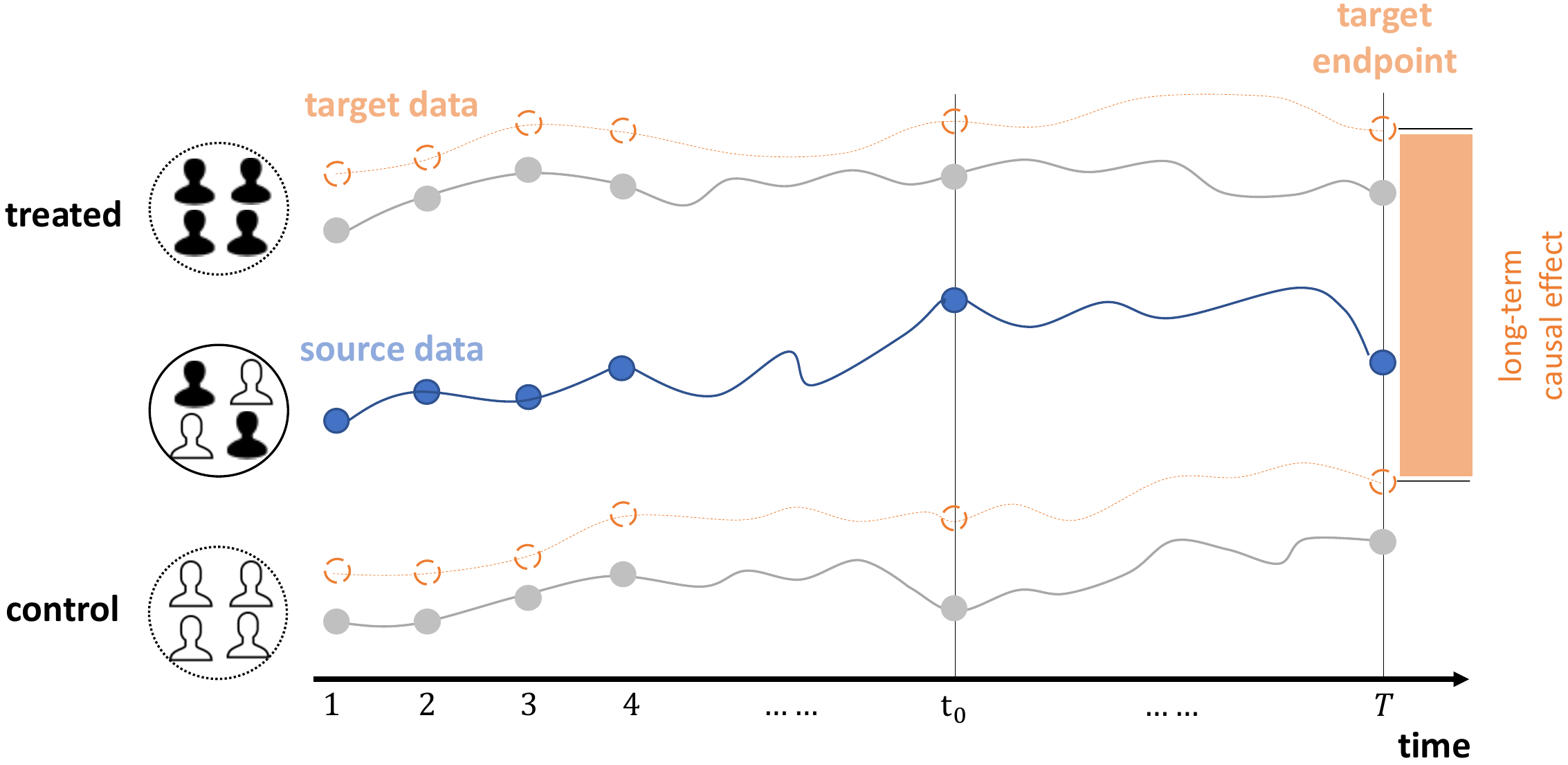}
  \caption{The two inferential tasks for estimating long-term causal effects in the target data (orange) given the source data (blue). First, inferring potential outcomes (grey and orange dots) across treatments at each timestep ($x$-axis), in particular, the top/bottom assignment where all units are in the treated/control group. Second, inferring across time, from $t=1$ to long-term $T$. What we seek, at the inferential target endpoint, is the difference between outcomes of the same units under treated and control in the target data (the orange rectangle at $T$). Note that we do not assume the short-term outcomes in the target data are observed.}
  \label{problem}
\end{figure}

The long-term effect of a treatment is defined as the difference between the primary outcomes of the same units under both treated and control. For example, the difference between the annual ad clicks from the same group of users exposed to more (treated) and fewer (control) ads. What is available in observational studies, however, is the \textit{source data} that typically consists of a sequence of \textit{short-term outcomes}, e.g., the first five-month ad clicks, and primary outcomes under one treatment assignment, e.g., the historical annual ad clicks from users who were exposed to fewer ads. We are interested in estimating the long-term effect in the \textit{target data}\footnote{The ``source'' and ``target'' data are referred to the ``observational'' and ``experimental'' data in some literature (e.g., \cite{athey2019surrogate}). We adopt the new terms to avoid the potential confusions with the ''observational data'' used in this work.}, where primary outcomes or even short-term outcomes cannot be observed (e.g., a short-term exposure to more ads may not be applicable for some online companies). Therefore, for long-term effect estimation, we generally need to perform two inferential tasks simultaneously as illustrated in Fig. \ref{problem}:
\begin{itemize}[leftmargin=*]
    \item[--] \textit{to infer outcomes across possible treatment assignments,} and 
    \item[--] \textit{to infer primary outcomes from the observational data\footnote{We term the source data and the short-term outcomes in the target data together as observational data. While required in some existing work, the short-term outcomes in the target data are not required in our work.}}.
\end{itemize}
The first task is the ``fundamental problem of causal inference'' \cite{holland1986statistics,rubin2005causal} in observational studies as we cannot observe outcomes of the same units under both treated and control. The resulting imbalance between the treated and the control distributions can cause large estimation error and variance \cite{shalit2017estimating}. For example, more ads will be likely assigned to users who are more active online. In long-term causal inference, the error and variance can accumulate over time towards the inferential target point. Therefore, achieving more consistent long-term effect estimations requires to reduce the imbalance between different treatment groups.

The second task presents unique challenges of long-term causal inference as we do not observe the primary outcome in the target data. A common approach employs related short-term variable as the proxy of the primary outcome, termed as a ``surrogate'' \cite{prentice1989surrogate}.
For example, the fifth-month ad clicks is the surrogate of annual ad clicks. A direct use of these short-term proxies requires the stringent ``surrogacy assumption'' \cite{begg2000use}: primary outcome is independent of the treatment conditioning on the surrogate. Difficult to warrant in observational studies, the surrogacy assumption inevitably impedes the applications of established methods.
Another standard assumption is the \textit{unconfoundedness} \cite{rubin1974estimating} that conditions on invariant confounders. However, confounders evolve over time in long-term effect estimation, e.g., users' age, personal preferences. Furthermore, these confounders may not be measured at each timestep. It is, therefore, essential to control for the time-varying confounding factors inferred from observed outcomes. Lastly, conventional approaches (e.g., \cite{athey2019surrogate,krueger1999experimental,chetty2011does,chetty2016effects}) often assume the linear relationship between short- and long-term outcomes. However, the non-linear time dependencies among the \textit{sequential} outcomes are particularly useful for predicting the primary outcomes \cite{dietterich2002machine}. 

To address the aforementioned challenges in each inferential task, we propose a principled framework -- \underline{L}ong-\underline{T}erm \underline{E}ffect \underline{E}stimation (LTEE). LTEE aims to learn \textit{surrogate representations} in a latent space that 1) reduce the imbalance between the induced distributions of the treated and the control at each timestep; 2) circumvent the stringent surrogate conditions; 3) account for the time-varying confounders; and 4) capture the time-dependencies among the observed outcomes.
To this end, LTEE relies on weaker assumptions that impose on independence between the treatment and primary outcome by conditioning on time-varying surrogate representations together shaped by the treatment, context and observed outcomes.

We summarize our contributions as follows:
After formally defining the problem of estimating long-term causal effects, we propose a principled framework LTEE that connects theoretical results in causal inference with sequential models in machine learning to improve the estimation efficiency. Accordingly, we propose weaker assumptions to circumvent standard stringent assumptions.
Empirical evaluations\footnote{The implementation code is available at https://github.com/GitHubLuCheng/LTEE} on two semi-synthetic datasets manifest that our proposed method outperforms the state-of-the-art even with no access to the short-term outcomes in the target data.
\section{Problem Definition}
We consider a setting with two data populations: source data $O=\{(\mat{x}_i, w_i, \mat{y}_i)\}$ with $N_O$ units and target data $E=\{(\mat{x}_i, w_i)\}$ with $N_E$ units. $P_i \in \{O,E\}$ denotes a binary indicator for the group to which the $i$-th unit belongs. For each unit in $O$, we can observe its context $\mat{x}_i$, the assigned treatment, $w_i\in\{0,1\}$, and a sequence of observed outcomes, denoted as $\mat{y}_i=<y_{i1}, y_{i2},...,y_{it_0},y_{iT}>$ where $y_{it}$ is the outcome observed at $t$. Short-term outcomes are measured at each discrete timestep $1,2,...,t_0$ and primary outcome at $T$.
We aim to estimate the long-term causal effect at $T> t_0$ in the target data $E$, including units' contexts and the treatment assignments. The key is to infer the primary outcome in $E$ that determines the long-term causal effect. $E$ might also include short-term outcomes. The problem of long-term effect estimation can be defined as 
\begin{definition}[\textbf{Long-Term Causal Effect Estimation}]
Given source data $O$ and target data $E$, we are interested in estimating the long-term average treatment effect (ATE) at $T$ in $E$, i.e., $\tau_T|E$:
\begin{equation}
    \tau_T|E=\mathbb{E}[\tau_{iT}]=\mathbb{E}[y_{iT}^1-y_{iT}^0|P_i=E]=\frac{1}{N_E}\sum_{i=1}^{N_E}(y_{iT}^1-y_{iT}^0).
\end{equation}
\end{definition}
\noindent Let $M^{W}$ be the surrogate variable under treatment $W$ (e.g., $Y_{t_0}^1$), conventional methods assume the independence between the primary outcome and the treatment given $M^{W}$, formally
\begin{assumption}[Surrogacy Assumption \cite{begg2000use}] \ \\
\begin{equation}
    Y^1_{T},Y^0_{T}\indep W|M^1,M^0.
\end{equation}
\end{assumption}
\noindent A major task of this work is to relax the surrogacy assumption.\\
\noindent\textbf{A Real-World Example.} At Google\footnote{http://google.com/}, and in Google Search Ads in particular, scientists have long recognized that optimizing for short-term revenue can be detrimental in the long-term due to a specific user learning effect: ads blindness and sightedness, i.e., users changing the possibility of clicking on or interacting with ads based on their prior experience with ads \cite{hohnhold2015focusing}. Suppose Google Search Ads is interested in the long-term effects of how ads are matched to queries on the annual revenue on two search engines. Google scientists first collected data from one search engine including whether to assign the change to a user, her/his attributes, the first five-month revenue and the annual revenue. The goal is to estimate the effect of the changes of ads-queries matching on annual revenue on the other search engine where the assigned treatment and users' attributes are observed, i.e., the target data, by leveraging the data collected from the first search engine, i.e., the source data. An implicit assumption is that the conditional distributions of the annual revenue in the two search engines are identical. Conventional methods using the fifth-month revenue to approximate the annual revenue assume that the annual revenue is independent of the changes given the fifth-month revenue. 
\begin{figure*}
\center
  \includegraphics[width=.9\textwidth]{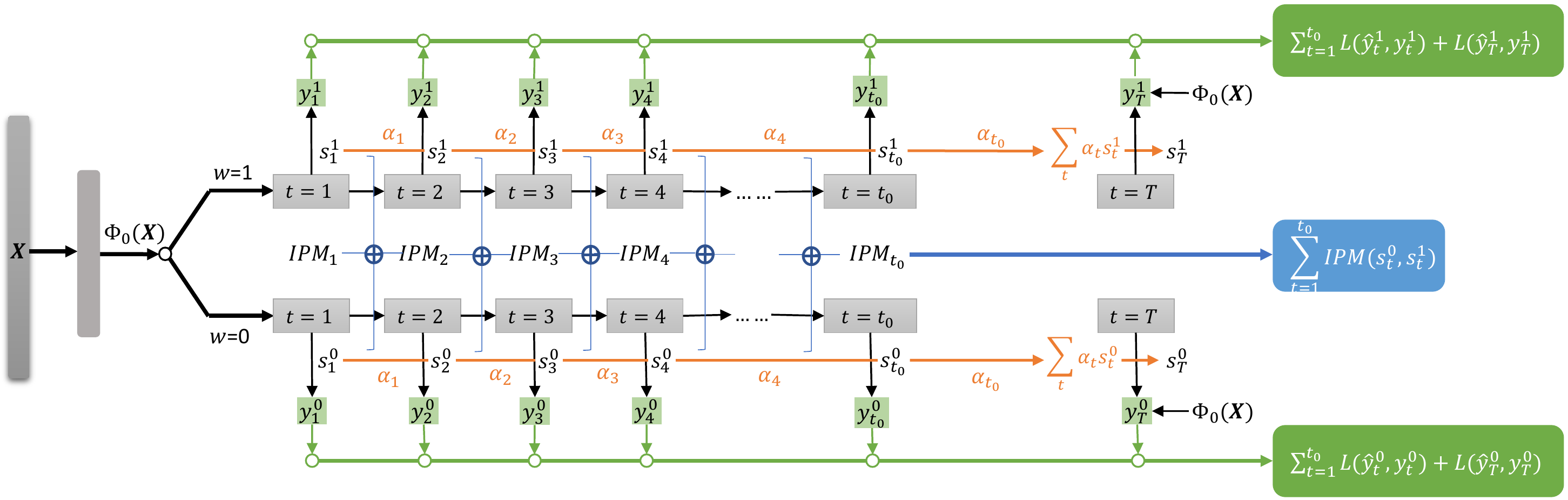}
  \caption{Illustration of the LTEE framework. LTEE first transforms the context $\mat{X}$ into a latent space and obtains the latent representation $\Phi_0(\mat{X})$. It then builds double-headed RNNs where each head represents an RNN specifically trained for the treated ($w=1$) or the control ($w=0$) group. At each timestep $t\in\{1,2,...,t_0\}$, the RNN outputs the surrogate representation (e.g., $s_1^1$ in the treated group) to perform two subtasks simultaneously: (1) predicting the outcome observed at $t$ and (2) enforcing the similarity between distributions of surrogate representations in the treated and the control groups. At long-term $T$, we can observe the outcome in the source data (but not target data), and $s_T^w$ is the sum of all previous surrogate representations weighted by the attention mechanism in \cite{bahdanau2014neural}. Consequently, the objective function of LTEE consists of the sum of the outcome prediction errors within each treatment assignment and imbalance errors (IPM) across treatment assignments at each timestep. Note that LTEE only uses the source data ($O$) for training.}
  \label{framework}
\end{figure*}
\section{LTEE: The Proposed Framework}
LTEE bridges the concept of surrogacy and sequential models with deep learning to circumvent the stringent surrogacy assumption. The proposed assumption conditions on \textit{surrogate representations} that i) redress the imbalance between distributions of the treated and control groups, ii) account for time-varying confounders, and iii) exploit non-linear structural information among observed outcomes. However, when the dimension of surrogate representation is high, risks of losing the influence of the treatment on outcome prediction can be large. Therefore, LTEE is designed to have \textit{double heads} that separate the learning of surrogate representations under treated from those under control.
At each timestep, LTEE simultaneously minimizes the outcome prediction loss and the distance between the treated and the control distributions induced by the surrogate representations.
The final surrogate representation of the primary outcome is the weighted sum of surrogate representations encoded at previous timesteps. 
The overview of the proposed framework LTEE can be seen in Fig. \ref{framework}.
\subsection{RNN with Double Heads}
Recurrent Neural Network (RNN) \cite{chen1996comparative} is one of the most common sequence models and has been successfully applied in numerous tasks such as natural language processing and time series forecasting.
Our task is inherently relevant to time series forecasting as both seek to predict outcomes in the future. Specifically, we feed the context and treatment assignment into a sequence model and it then generates the output which is fed back to the model as a new input.
At each timestep (state) $t$ for unit $i$, the model predicts the outcome $y_{it}$
based on the treatment $w_i$ and representation $\mat{s}_{it}\in \mathbb{R}^D$ output from each hidden state.
Suppose we observe $t_0$ short-term outcomes and the primary outcome at $T$ in the source data $O$, a straightforward approach to estimating the ATE at $T$ is by \textit{direct modelling} \cite{johansson2016learning}: given $N_O$ samples $\{\mat{x}_i, w_i, \mat{y}_i\}_{i=1}^{N_O}$, we have $y_{it}\sim p(Y_t^1|\mat{x}_i)$ if $w_i=1$ and $y_{it}\sim p(Y_t^0|\mat{x}_i)$ if $w_i=0$. This consistency assumption implies that the treatment assignment determines which potential outcome to see. We can then train hypothesis $h_t: \Phi_t(\mathcal{X})\times \mathcal{W}\rightarrow \mathcal{Y}, t\in \{1,2,...,t_0,T\}$ such that $h_t(\mat{s}_{it},w_i)\approx y_{it}$, where $\mat{s}_{it}=\Phi_t(\mat{x}_i)$. The transductive individual treatment effect (ITE) at long term $T$ is then estimated as
\begin{equation}
    \hat{ITE}_T(\mat{x}_i)=h_T(\mat{s}_{iT},1)-h_T(\mat{s}_{iT},0).
\end{equation}
We can estimate long-term ATE by taking the average of $\hat{ITE}_T(\mat{x}_i)$ over all units. However, in this direct modelling method, influence of the treatment on training hypotheses $\{h_t\}$ might vanish over time especially when the dimension of $\mat{s}_{it}$ is high \cite{shalit2017estimating}. A simple concatenation of the treatment and latent representation as the input at each timestep cannot effectively reduce the risks. To combat this, we propose to build RNN for each treatment group separately.
Specifically, let $\mathcal{H}^1=\{h^1_t\}$ and $\mathcal{H}^0=\{h^0_t\}$ be the set of hypotheses in the treated and control groups, we parameterize $\mathcal{H}^1$ and $\mathcal{H}^0$ as two separate ``heads'' of the joint RNNs, the former used to estimate the outcome under treated and the latter under control, as shown in Fig. \ref{framework}. The separate heads indicate that statistical power is shared in the hidden layers of the joint RNNs, while the effect of treatment is retained in the separate heads \cite{shalit2017estimating}. Note that each unit is only used to update the RNN corresponding to its observed treatment. For example, an observation ($\mat{x}_i,w_i=1,\mat{y}_i$) is only used to update $\mathcal{H}^1$. In the double-headed RNN, ITE of $i$ at $T$ is now estimated as
\begin{equation}
    \hat{ITE}_T(\mat{x}_i)=h^1_T(\mat{s}^1_{iT},1)-h^0_T(\mat{s}^0_{iT},0).
\end{equation}
In practice, we employ Gated Recurrent Units (GRU) based RNN (GRU-RNN) \cite{bahdanau2014neural} to encode the sequence of observed outcomes. GRU uses update gate and reset gate to solve the vanishing gradient problem of a standard RNN. Note that our work is flexible to other RNN architectures, such as long-short term memory (LSTM) \cite{hochreiter1997long}, which are left to explore in the future.
\subsection{LTEE Assumptions}
The common approach for estimating long-term causal effect directly uses short-term outcomes as ``statistical surrogate'' \cite{prentice1989surrogate} and thereby relies on the stringent ``surrogacy assumption'' \cite{begg2000use}.
In addition, the standard unconfoundedness assumption \cite{rubin1974estimating,guo2018survey} conditional on time invariant confounders $\mat{x}$ is often violated given the dynamic nature of long-term causal inference. For example, users' age and preferences change over time. 

In this study, we seek weaker assumptions by a revisit of the short-term outcomes from a machine learning perspective. Specifically, we view the observed outcomes as a sequence of ``labels'' of unit $i$ at each timestep $t$. Then the samples in $O$ are fed into the double-headed RNNs that predict outcome $y^{w_i}_{it}$ using the treatment and hidden state $\mat{s}^{w_i}_{it}$. We refer to $\mat{s}^{w_i}_{it}$ as the \textit{surrogate representation} at $t$ under treatment $w_i$. The direct result is that we can circumvent the stringent surrogacy assumption by conditioning on latent surrogate representations with following features: First, $\mat{s}^{w_i}_{it}$ gets updated over time by receiving supervised signals from $\mat{y}$, therefore, can effectively account for time-varying confounding factors; Second, by incorporating the temporal order of observed outcomes, $\mat{s}^{w_i}_{it}$ leverages the strength of sequential models to keep track of the long-term dependencies among outcomes.

There are two primary assumptions that together allow LTEE to exploit source data $O$ to estimate the effect on the primary outcome in target data $E$. The first assumption is the \textit{temporal unconfoundedness}, assuming that all the factors determining treatment assignment at each timestep are included in the corresponding surrogate representations. The second assumption is \textit{comparability}, assuming that whether a sample belongs to $E$ or $O$ depends only on other variables such as context and treatment. It ensures the eligibility of inferring relationships in the target data from the source data. 
\begin{assumption}[Temporal Unconfoundedness]\ \\
(i) Treatment is independent of potential outcomes at each timestep given the corresponding surrogate representations. Formally,
\begin{displaymath}
     (y_{it}^0, y_{it}^1) \; \indep \; w_i| \; \mat{s}_{it}^0, \mat{s}_{it}^1 \quad \forall t.
\end{displaymath}
(ii) The probability of a unit being assigned to a particular treatment given its context is positive. Or,
\begin{displaymath}
    0<e(\mat{x}_i)<1 \quad \forall \mat{x}_i \in \mathcal{X},
\end{displaymath}
\end{assumption}
\noindent where $e(\mat{x}_i)=p(w_i=1|\mat{x}_i)$, or the propensity score \cite{rubin1974estimating}. Asm.2 indicates that the surrogate representations fully capture the causal links between the treatment and the outcomes. Its benefits are two-fold. First, in contrast to the conventional unconfoundedness assumption, Asm.2 implies that treatment effect on the outcome at each timestep can be estimated by adjusting for the time-varying confounders $\mat{s}^{w_i}_{it}$, if the outcomes are measured. Note that the conventional unconfoundeness entails all the potential outcomes are jointly independent of the treatment whereas Asm.2 imposes on independence between treatment and potential outcomes at each timestep. Second, Asm. 2 enables LTEE to circumvent the surrogacy assumption by replacing the conditioned short-term proxies with the surrogate representation $\mat{s}^{w_i}_{it}$ together shaped by the treatment, contexts, and observed outcomes. 
\begin{assumption}[Comparability (an extension from Asm.3 in \cite{athey2019surrogate})] The conditional distribution $p(y^{w_i}_{it}|(\mat{s}^{w_i}_{it},w_i, \mat{x}_i,P_i=O))$ in the source data is identical to that in the target data $p(y^{w_i}_{it}|(\mat{s}^{w_i}_{it},w_i,\mat{x}_i,P_i=E))$:
\begin{displaymath}
y^{w_i}_{it}|\mat{s}^{w_i}_{it}, w_i, \mat{x}_i, P_i=O \sim y^{w_i}_{it}|\mat{s}^{w_i}_{it}, w_i , \mat{x}_i, P_i=E \quad \forall t.
\end{displaymath}
\end{assumption}
\noindent The comparability assumption places restrictions on how the relationship between $Y$ and other variables in $O$ compares to those in $E$. This assumption validates the eligibility of using source data to infer primary outcome in the target data.
\subsection{Long-Term Causal Effect Estimation}
With the LTEE assumptions, this section first discusses how to use the introduced surrogate representations to infer the potential outcomes across time. It then illustrates a balancing strategy that minimizes the distance between the treated and control distributions over the surrogate representations to reduce the imbalance.
\subsubsection{Inferring outcomes across time}
For unit $i$, we first transform the context $\mat{x}_i$ into a latent space: $\mat{x}_i\rightarrow \Phi_0(\mat{x}_i)$. The resulting dense vector $\Phi_0(\mat{x}_i)$ is then fed into the double-headed RNNs. At timestep $t$ in RNN w.r.t. treatment $w_i$, the surrogate representation $\mat{s}^{w_i}_{it}$, along with $w_i$ is used to predict the outcome $y^{w_i}_{it}$ under the hypothesis $h^{w_i}_t$. We define the loss for predicting short-term outcomes over the source data $O$ as 
\begin{equation}
    \ell_1=\frac{1}{N_O\cdot t_0}\sum_{i=1}^{N_O}\sum_{t=1}^{t_0}L(h_t^{w_i}(\mat{s}^{w_i}_{it},w_i),y^{w_i}_{it})+\lambda \cdot \mathcal{R}(\mathcal{H}) \quad w_i\in\{0,1\},
\end{equation}
where $\mathcal{H}=\mathcal{H}^0\cup \mathcal{H}^1$, $L$ is a function measuring the prediction error, $\mathcal{R}$ is a model complexity term weighted by the hyperparameter $\lambda$. $L$ can be mean squared error or cross-entropy loss depending on whether the outcome is continuous or categorical. Further, the short-term outcomes may not be equally important to the primary outcome. For example, the seventh-month ad clicks is typically more informative than the first-month's w.r.t. estimating the effect of new ads on the annual ad clicks. Consequently, we leverage the attention mechanism \cite{bahdanau2014neural} to construct the final surrogate representation of the primary outcome. We first feed the surrogate representation to a dense layer and get the hidden state of $\mat{s}^{w_i}_{it}$:
\begin{equation}
     \Tilde{\mat{s}}^{w_i}_{it}=\tanh(\mat{A}\mat{s}^{w_i}_{it} + \mat{b}),
\end{equation}
where $\mat{A}$ is the connection weight matrix and $\mat{b}$ is the bias. Suppose that a latent vector $\mat{u}$ captures all the structural information between outcomes, we calculate the similarity between $\mat{u}$ and $\Tilde{\mat{s}}^{w_i}_{it}$ as follows:
\begin{equation}
    \alpha_{it}=\frac{\exp(\Tilde{\mat{s}}^{w_i\intercal}_{it} \mat{u})}{\sum_t\exp(\Tilde{\mat{s}}^{w_i\intercal}_{it} \mat{u})}.
\end{equation}
Here, $\alpha_{it}$ is a normalized importance weight for $\mat{s}^{w_i}_{it}$. The surrogate representation of the primary outcome is the sum of the weighted surrogate representations from previous timesteps:
\begin{gather}
    \mat{s}^{w_i}_{iT}=\sum_t \alpha_{it}\mat{s}^{w_i}_{it}.
\end{gather}
Finally, the prediction error of the primary outcome in the source data $O$ can be defined as 
\begin{equation}
    \ell_2=\frac{1}{N_O}\sum_{i=1}^{N_O}L(h_T^{w_i}(\Phi_0(\mat{x}_i), \mat{s}^{w_i}_{iT}, w_i),y^{w_i}_{iT}) \quad w_i\in\{0,1\},
\end{equation}
Here, we also include $\Phi_0(\mat{x}_i)$ to retain information of unit's context.
\subsubsection{Balancing Surrogate Representation}
In observational studies, units assigned to the treated can be very different from those assigned to the control. For example, active online users are likely to receive more ads compared to non-active users. The bias induced by imbalance across the treatment assignments can cause large estimation error and variance in long-term effect estimation. Consequently, LTEE simultaneously minimizes the prediction error of outcomes and the imbalance between the distributions of the treated and control groups induced by the surrogate representations.

Here, we employ the Integral Probability Metric (IPM) \cite{muller1997integral} measure of distance between the two distributions $p(\mat{S}|W=0)$ and $p(\mat{S}|W=1)$. IPM is a class of distance metrics between probability distributions \cite{sriperumbudur2012empirical}. In practice, we use the Wasserstein-1 (W-1) distance \cite{cuturi2014fast} because it has consistent estimators which can be efficiently computed in the finite sample case \cite{sriperumbudur2012empirical}. Let $P_t=p(\mat{s}_{it}^1)$ and $Q_t=p(\mat{s}_{it}^0)$ be the empirical distributions of surrogate representations in the treated and control groups at $t$, the W-1 distance at each timestep is defined as \cite{cuturi2014fast}
\begin{equation}
    \ell_3=\text{W-1}(P_t, Q_t):=\inf_{k\in \mathcal{K}_t}\int_{\mat{s}\in\{\mat{s}_{it}^1\}_i }\|k(\mat{s})-\mat{s}\|P_t(\mat{s})d\mat{s},
\end{equation}
where $\mathcal{K}_t$ denotes the set of Lipschitz-1 functions that seek to transform the surrogate representation distribution of the treated ($P_t(\mat{s})$) to that of the control ($Q_t(\mat{s})$) at $t$, i.e. $Q_t(k(\mat{s}))=P_t(\mat{s})$ for $k\in\mathcal{K}_t$. At each timestep $t$, LTEE predicts outcomes for the treatment and control groups on top of the ``balanced'' surrogate representations.
To this end, LTEE jointly minimizes the prediction error of outcomes and the W-1 distance between the control and treated distributions induced by the surrogate representations.
The final objective function of LTEE can be defined as
\small
\begin{align}
\mathcal{L}&=\frac{1}{N_O}\sum_{i=1}^{N_O}\big(\frac{1}{t_0}\sum_{t=1}^{t_0}L(h_t^{w_i}(\mat{s}^{w_i}_{it},w_i),y^{w_i}_{it})+L(h_T^{w_i}(\Phi_0(\mat{x}_i), \mat{s}^{w_i}_{iT},w_i),y^{w_i}_{iT})\big) \nonumber \\
&+\gamma\cdot\sum_{t=1}^{t_0}\text{W-1}(\{\mat{s}_{it}^1\},\{\mat{s}_{it}^0\}) +\lambda \cdot \mathcal{R}(\mathcal{H}) \quad w_i\in\{0,1\},
\label{loss}
\end{align}
\normalsize 
where $\gamma$ controls the contributions of the balancing term. Eq. \ref{loss} can be viewed as learning functions that predict primary outcomes under a constraint that encourages better generalization across the treated and control populations at each timestep. We train LTEE on $O$ by minimizing Eq. \ref{loss} using stochastic gradient descent, where we backpropagate the error through the outcome prediction and representation learning. Both the prediction error and penalty term W-1($\cdot,\cdot$) are computed for one mini-batch at a time.

\noindent\textbf{Long-Term Effect Estimation in Target Data.} After we train LTEE on the source data, we feed the target data $\{(\mat{x}_i,w_i)\}$ into LTEE and predict the primary outcomes under two treatment assignments $\hat{y}_{iT}^1$ and $\hat{y}_{iT}^0$. ATE at $T$ in $E$ is then estimated as 
\begin{equation}
    \hat{\tau}_T|E=\frac{1}{N_E}\sum_{i=1}^{N_E}(\hat{y}_{iT}^1-\hat{y}_{iT}^0).
\end{equation}
\section{Experiments}
In this section, we begin by presenting the experimental design including the data simulation process and a summary of established methods used for performance comparisons.
Then, we illustrate experimental results by evaluating the gains from using surrogate representations in terms of precision relative to existing methods. Essentially, we evaluate the proposed framework LTEE with a focus on four research questions: \\
\noindent\textbf{RQ1.} How does LTEE fare against established methods for estimating long term effects? In particular, we examine the performance of LTEE and existing methods 1a) with varied $t_0$ and fixed $T$; and 1b) with fixed $t_0$ and varied $T$.\\ 
\noindent\textbf{RQ2.} What is the effect of imposing the imbalance regularization on the surrogate representations?\\
\noindent\textbf{RQ3.} What is the effect of the ``double heads'' in LTEE?\\
\noindent\textbf{RQ4.} What is the effect of the non-linearity in LTEE?
\subsection{Experimental Design}
We evaluate the models on two benchmark datasets \cite{cheng2019practical} for causal effect estimation--Infant Health and Development Program (IHDP)~\cite{hill2011bayesian} and News~\cite{johansson2016learning}. For both datasets, we use the original context, treatment assignment and synthetic outcomes (more details later) that reflect the sequential observations in long-term causal inference.
The IHDP dataset is from a real-world randomized experiment that studies the effect of high-quality child care and home visits on the children's cognitive test scores.
To mimic the real-world scenario, the imbalance between the treated and control units has been artificially introduced by removing a subset of the treated population \cite{hill2011bayesian}. In total, IHDP consists of 747 units (139 treated, 608 controlled), each represented by 25 features indicating attributes of the child and her/his mother. Please refer to ~\cite{hill2011bayesian} for more details. 

The second dataset, News \cite{johansson2016learning}, simulates readers experience exposed to multiple news items.
Each item is consumed either on a mobile device or on desktop. Units are different news items represented by word counts $\mat{x}_i \in \mathbb{N}^V$, where $V$ is the number of words in the corpus. Outcome $y_i\in \mathbb{R}$ is the reader experience of $\mat{x}_i$ modeled via a topic model trained on a large set of documents. Two centroids are defined in the topic space $\mat{z}_1^c$ (mobile) and $\mat{z}_0^c$ (desktop). Let $k$ be the number of topics and $\mat{z}(\mat{x}_i)\in \mathbb{R}^k$ be the topic distribution of news item $\mat{x}_i$, the outcome of news item $\mat{x}_i$ on device $t$ is determined by the similarity between $\mat{z}(\mat{x}_i)$ and $\mat{z}_t^c$: $y^{w_i}_i=C\big(\mat{z}(\mat{x}_i)^\intercal\mat{z}_0^c+w_i\cdot\mat{z}(\mat{x}_i)^\intercal\mat{z}_1^c\big)+\epsilon$, where $C$ is a scaling factor and $\epsilon \sim \mathcal{N}(0,1)$ denotes the noise. 
The treatment $w_i\in\{0,1\}$ indicates the viewing device, desktop ($w_i=0$) or mobile ($w_i=1$). It is simulated under the assumption that the assignment is biased towards the device preferred for a specific item. News consists of 5,000 news items and each news item is represented by 3,477 word counts. For more details, see \cite{johansson2016learning}.

\subsubsection{Simulating Outcomes.} We further simulate the short-term and primary outcomes for both datasets to know the ground truth for the causal effect at each timestep. 
We assume that $y_{it}$ observed at $t$ is determined by three factors: context $\mat{x}_i$, treatment $w_i$ and all the outcomes observed before $t$.
To illustrate, in the example of Google Search Ads, the annual revenue of a search engine depends on users' attributes, whether to change the ads-queries matching for a user and the first five-month revenue. While we recognize that the primary outcome is not necessarily always related to short-term outcomes, there are many cases where this is possible. To this end, our outcome simulation methods mainly consist of two components: the first component follows the original outcome generating processes introduced in \cite{hill2011bayesian} and \cite{johansson2016learning}; the second component further imposes the causal influence of historical outcomes on the current outcome. For simplicity, we use the average of all the historical outcomes as the second component. For IHDP dataset, the potential outcome of unit $i$ at $t$ is generated as follows:
\begin{equation}
    y^{w_i}_{it}(\mat{x}_i) \sim \begin{cases}
      \mathcal{N}(\mat{x}_i\beta,1)+\frac{C_1}{t-1}\sum_{j=1}^{t-1}y^{w_i}_{ij}, & w_i=0 \\
      \mathcal{N}(\mat{x}_i\beta+4,1)+\frac{C_1}{t-1}\sum_{j=1}^{t-1}y^{w_i}_{ij}, & w_i=1,
    \end{cases} 
\end{equation}
where the coefficients $\beta$ are randomly sampled from $(0,1,2,3,4)$ with probabilities $(0.5,0.2,0.15,0.1,0.05)$, as defined in \cite{hill2011bayesian}. $C_1$ is a scaling factor. Similarly, for News dataset, we incorporate the outcome simulation process used in \cite{johansson2016learning} with the average of all the historical outcomes:
\begin{equation}
    y^{w_i}_{it}(\mat{x}_i) = C\big(\mat{z}(\mat{x}_i)^\intercal\mat{z}_0^c+w_i\cdot\mat{z}(\mat{x}_i)^\intercal\mat{z}_1^c\big)+\frac{C_2}{t-1}\sum_{j=1}^{t-1}y^{w_i}_{ij}+\epsilon,
\end{equation}
where $C_2$ is a constant scaling the effect of historical outcomes.
\begin{table}[]
\small
\setlength{\tabcolsep}{.8mm} 
    \begin{tabular}{|c|c|c|c|c|c|c|c|c|}\hline
        Models &  SInd& Na\"ive-I&Na\"ive-II&Na\"ive-III&CFO&TARNet&Inter&LTEE\\ \hline
        $O_{ST}$ &  \cmark&\xmark&\xmark&\xmark&\cmark&\cmark&\cmark&\cmark\\ \hline
        $O_{LT}$ &  \cmark&\cmark & \xmark&\xmark &\xmark & \xmark& \xmark& \cmark\\ \hline
        $E_{ST}$ & \cmark&\xmark&\cmark&\cmark & \cmark& \cmark&\cmark& \xmark \\ \hline
        $E_{LT}$ & \xmark &\xmark &\xmark &\xmark &\xmark &\xmark &\xmark &\xmark \\ \hline
    \end{tabular}
    \caption{Illustrations of different settings required in various models. There are four types of outcomes in total: short-term/primary outcome in $O$, denoted as $(O_{ST},  O_{LT})$; short-term/primary outcome in $E$, denoted as $(E_{ST},  E_{LT})$. \text{\cmark} indicates a model needs a certain type of outcome for training.}
    \label{setting}
\end{table}
\subsubsection{Baseline.} To answer the first question, we consider several baseline models including the-state-of-the-art method Surrogate Index \cite{athey2019surrogate} for long-term causal effect estimation, the Na\"ive methods and common causal effect estimation models.
We also include a method that handles the problem of missing value on time-series data.
While LTEE only has access to short-term and primary outcomes in the source data, some of the baselines also need the short-term outcomes in the target data for training. We summarize different data settings required for LTEE and baselines in Table \ref{setting}. Each baseline is briefly described as follows:
\begin{itemize}[leftmargin=*]
    \item \textbf{Surrogate Index (SInd) \cite{athey2019surrogate}.} SInd uses a ``surrogate index'' -- predictions from the Linear Regression model given the short-term outcomes and the context -- as the surrogate of the primary outcome. It assumes that the primary outcome is independent of the treatment conditional on the surrogate index. 
    \item \textbf{Na\"ive.} Following \cite{athey2019surrogate}, we consider several na\"ive methods that use short-term outcomes to predict the primary outcome. \textbf{Na\"ive-I} uses the outcome observed at $T$ in $O$ to approximate the primary outcome in $E$. Suppose we also observe the short-term outcomes in $E$, \textbf{Na\"ive-II} uses the outcome observed at $t_0$ in $E$ to approximate the primary outcome. \textbf{Na\"ive-III} uses the average of all the outcomes observed before and at $t_0$ in $E$ as the proxy. 
    \item \textbf{Causal Forest (CFO) \cite{wager2018estimation}.} CFO, extended from random forest algorithm, is one of the most popular methods for conventional ATE estimation. 
    It is composed of causal trees that estimate the effect at the leaves. 
    \item \textbf{TARNet \cite{shalit2017estimating}.} TARNet, a deep-learning-based method for conventional causl effect estimation, seeks to predict ITE by learning ``balanced'' representations such that the induced treated and control distributions look similar. 
    \item \textbf{Interpolate (Inter).} Inter tackles missing value on time series data via linear interpolation. It searches for a straight line that passes through the end points. 
\end{itemize}
We report the widely used evaluation metric--absolute error--in estimating long-term ATE for the target data, denoted as $\epsilon_{ATE}$. It is calculated based on the following formula:
\begin{equation}
    \epsilon_{ATE}=|\tau_T-\hat{\tau}_T|.
\end{equation}
CFO is implemented using the ``grf'' R-package \cite{tibshirani10grf} and TARNet is implemented using the open source code\footnote{https://github.com/clinicalml/cfrnet}. To adapt these two methods to long-term causal inference, we take the average of all the short-term outcomes as the outcome for CFO and TARNet. We split the simulated data such that 80\% is used as source data and the rest as target data under the comparability assumption. To avoid overfitting, we generate 10 repeated experiments for each dataset. We set the scaling parameters of the simulation processes as follows: $C$ is set to 50, $C_1$ and $C_2$ are set to 0.02 and 0.03, respectively. LTEE and TRANet are trained using Adam, with a small $l_2$ weight decay, $\lambda=10^{-6}$. $\gamma$ is set to $10^{-8}$ for IHDP and $10^{-10}$ for News based on the hyperparameter search detailed in Sec. 4.4. For TRANet, $\gamma$ is set to $10^{-6}$ for both datasets using same hyperparamter search method.

\subsection{Varying \texorpdfstring{$t_0$}{Lg} and Fixing \texorpdfstring{$T$}{Lg}}
\begin{figure}
\centering
\begin{subfigure}{.5\columnwidth}
\centering
  \includegraphics[width=.9\linewidth]{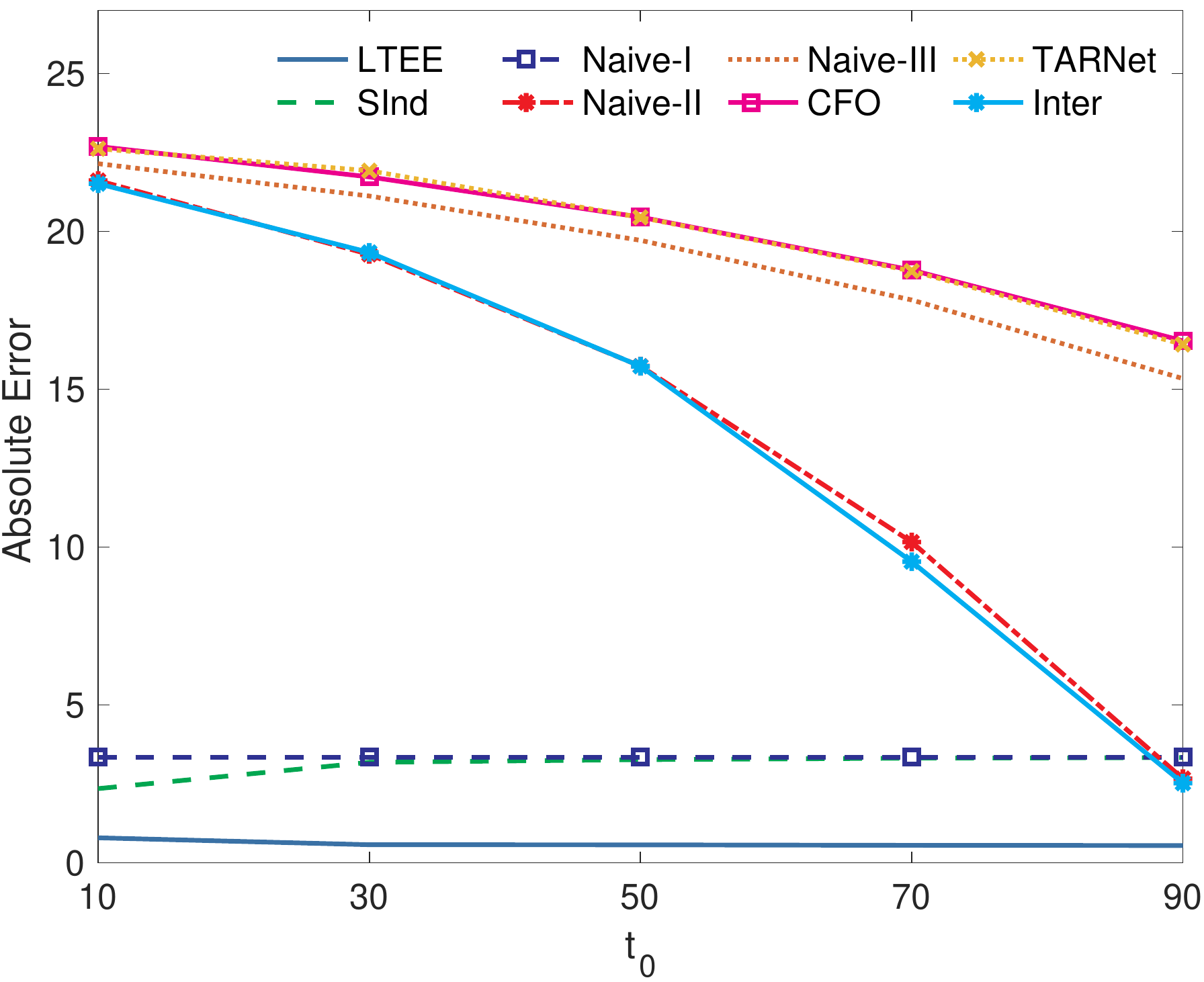}
  \caption{Results for IHDP data.}
\end{subfigure}%
\begin{subfigure}{.5\columnwidth}
\centering
  \includegraphics[width=.9\linewidth]{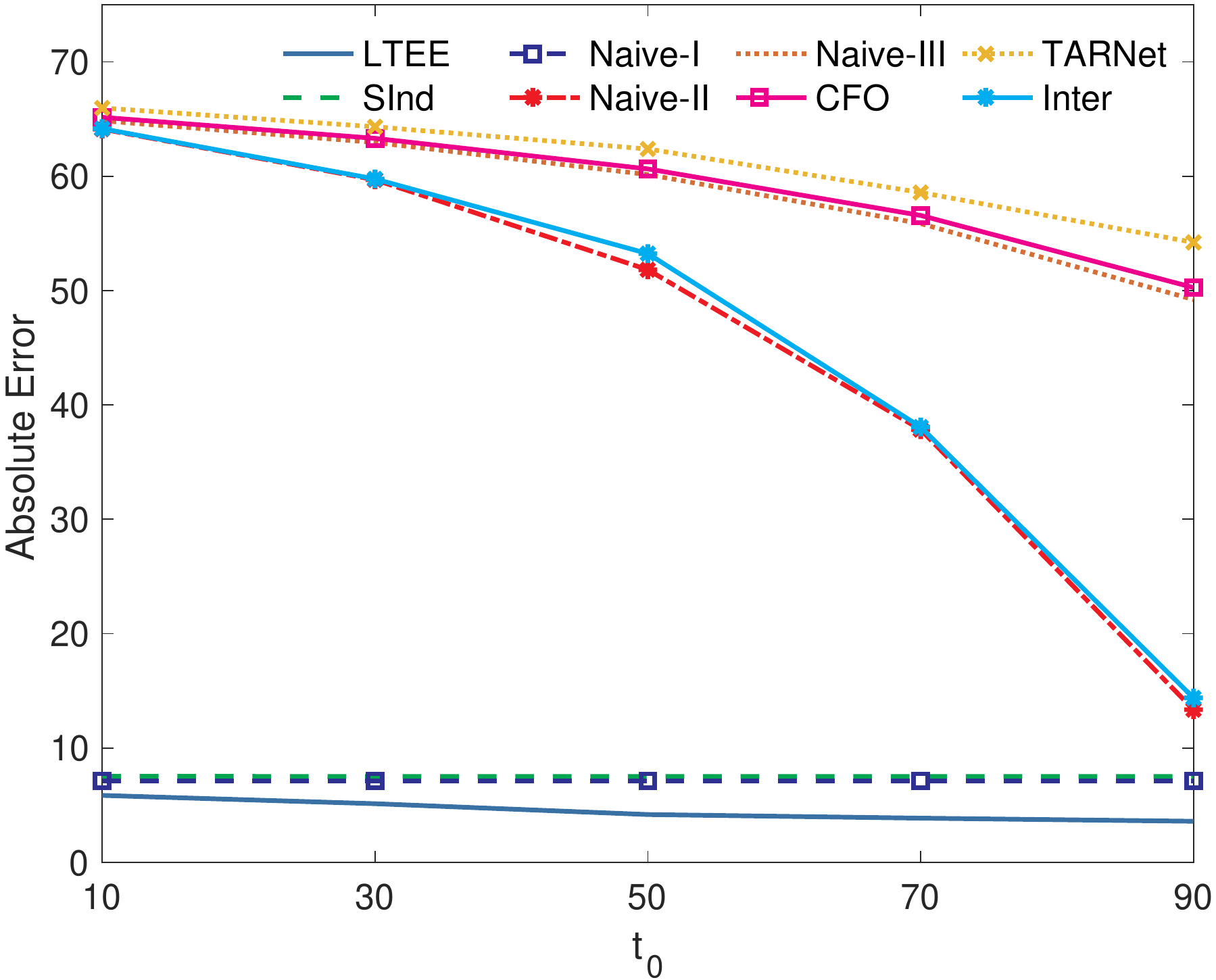}
  \caption{Results for News data.}
\end{subfigure}
\caption{Estimation error when varying $t_0$ and fixing $T$.}
 \label{t_0}
\end{figure}

We start with estimating long-term causal effect at $T$ with different $t_0$, i.e., how the number of short-term outcomes affects the models' performance in long-term ATE estimation.
We set $T=100$ and vary $t_0$ from 10 to 90 with a $20$ increment. The results for the IHDP and News datasets are presented in Fig. \ref{t_0}. We see that, in general, models (Na\"ive-III, TARNet, CFO) that take the average short-term outcomes in either target data or source data as the proxy of primary outcome show less competitive performance.
This is primarily because these models average out the important information embedded in the sequence of short-term outcomes.
Na\"ive-II and Inter perform equally well with different $t_0$ as they both rely on $y_{it_0}$ in the target data for the inference of primary outcome. These two baselines outperform the aforementioned three models and the improvement increases as $t_0$ becomes large. This observation suggests that historical outcomes may introduce undesired noise that exacerbates the performance if used inappropriately. This finding can be further supported by the slightly aggravated performance of SInd as $t_0$ increases.
We also observe that performance of most aforementioned models improves when more short-term outcomes are observed, especially for Na\"ive-II and Inter. LTEE, SInd and Na\"ive-I are more robust to changes in $t_0$.
Of particular interest is that SInd and Na\"ive-I present nearly identical results (explained in Sec. 4.3). 
%
%
%
\subsection{Varying \texorpdfstring{$T$}{Lg} and Fixing \texorpdfstring{$t_0$}{Lg}}
\begin{figure}
\centering
\begin{subfigure}{.5\columnwidth}
\centering
  \includegraphics[width=.9\linewidth]{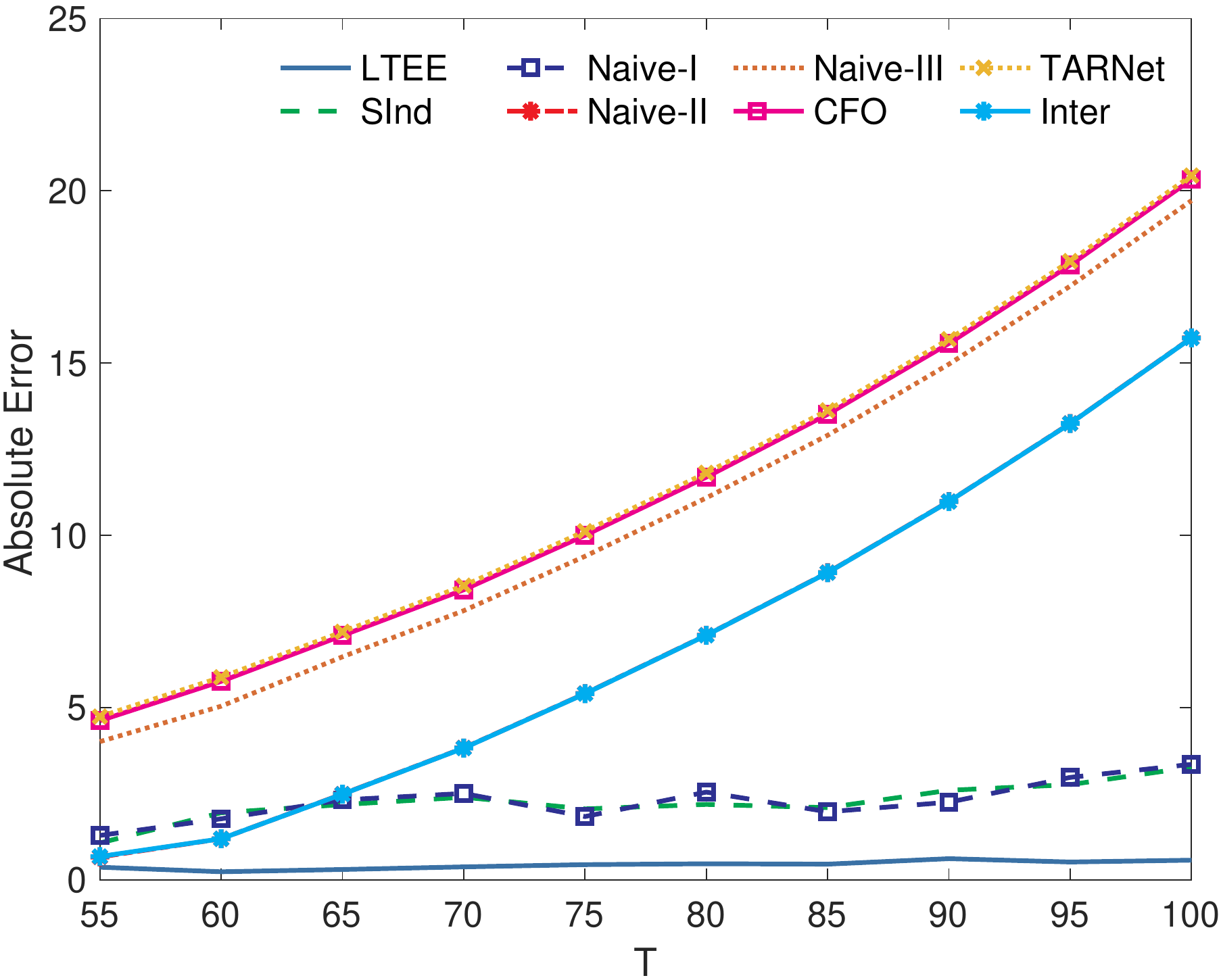}
  \caption{Results for IHDP data.}
\end{subfigure}%
\begin{subfigure}{.5\columnwidth}
\centering
  \includegraphics[width=.9\linewidth]{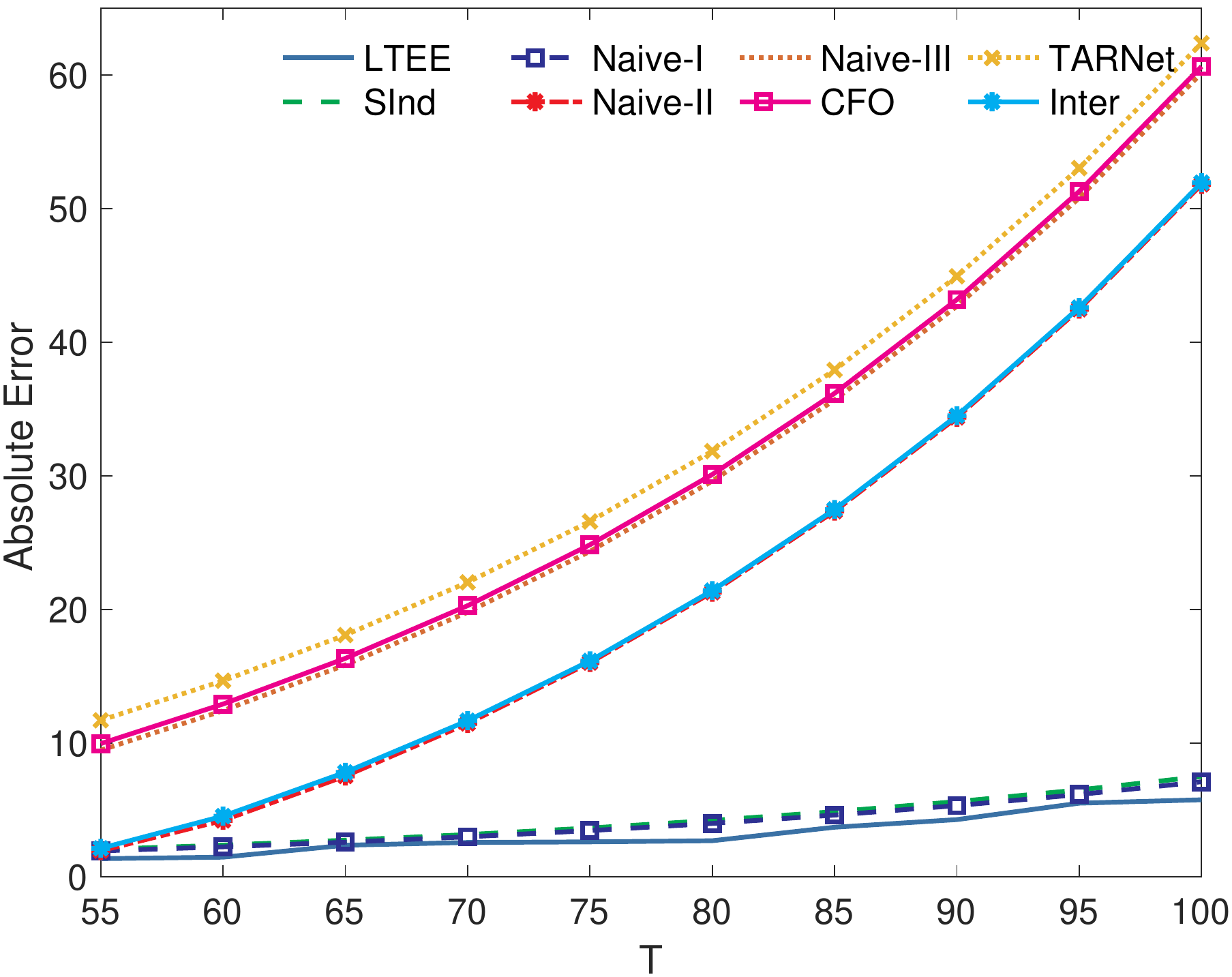}
  \caption{Results for News data.}
\end{subfigure}
\caption{Estimation error when varying $T$ and fixing $t_0$.}
 \label{T}
\end{figure}

We further evaluate LTEE's precision of estimating long-term effect at different $T$ with fixed number of short-term outcomes ($t_0$), i.e., how the performance changes when using same number of short-term outcomes to estimate primary outcomes after different time delays.
We set $t_0=50$ and vary $T$ from 55 to 100 with a 5 increment of the sequence.
The results for the IHDP and News datasets are presented in Fig. \ref{T}.
As expected, all models' performance exacerbates as $T$ becomes larger, i.e., there is a larger time delay between short-term and primary outcomes.
Methods that use the average of short-term outcomes as the proxy show the largest estimation error, followed by methods that rely on the outcomes observed at $t_0$. 
LTEE consistently outperforms all baselines with different $T$ in terms of estimation accuracy.
Additionally, SInd and Na\"ive-I perform equally well with different $T$, implying that under the linear-relationship assumption, including more observed short-term outcomes cannot provide extra useful information compared to the surrogate that uses the most recently observed outcome. 
\subsection{Ablation Study}
\subsubsection{Impact of Imbalance Regularization}
\begin{figure}
\centering
\begin{subfigure}{.5\columnwidth}
\centering
  \includegraphics[width=\linewidth]{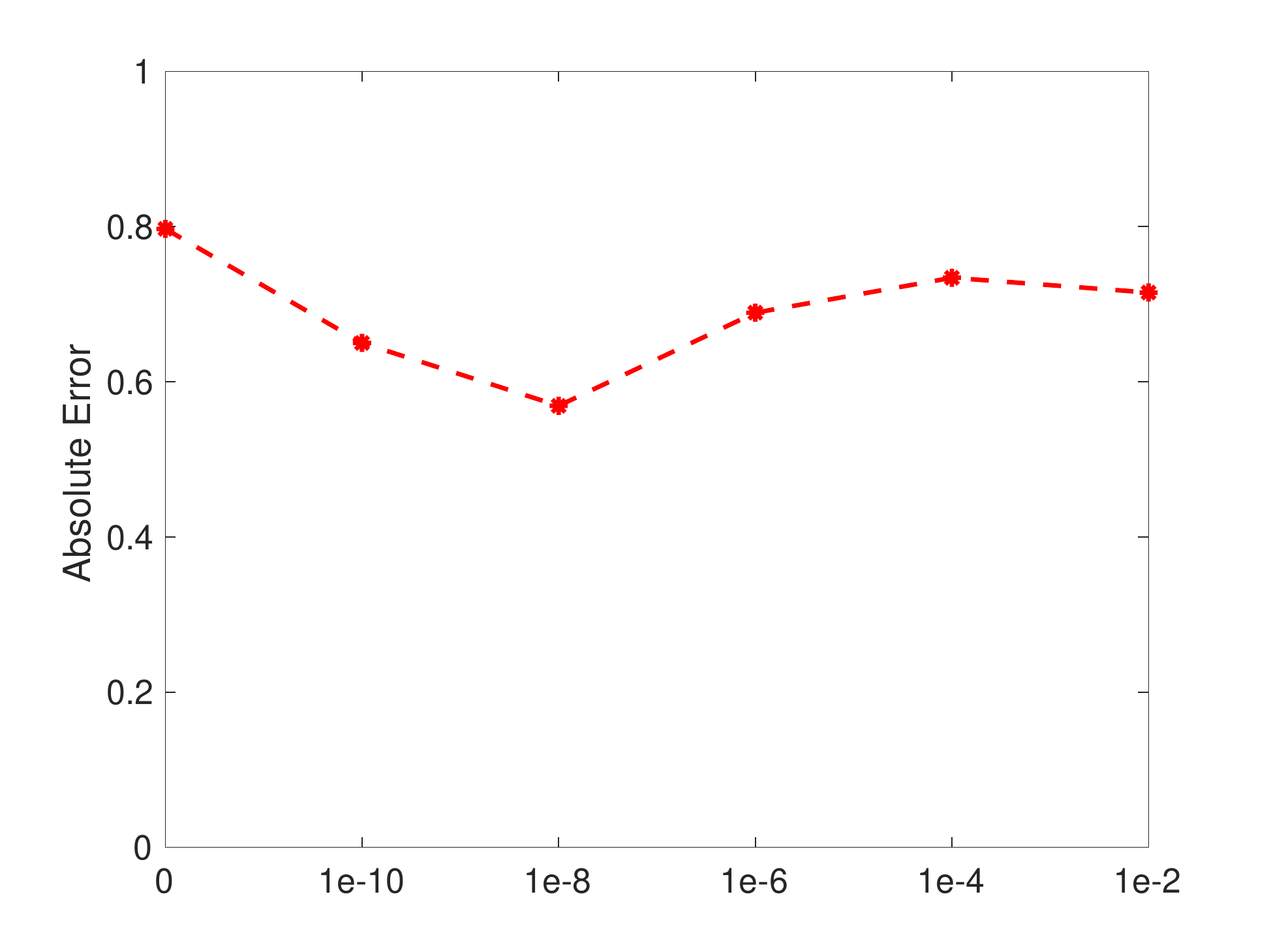}
  \caption{Results for IHDP data.}
\end{subfigure}%
\begin{subfigure}{.5\columnwidth}
\centering
  \includegraphics[width=\linewidth]{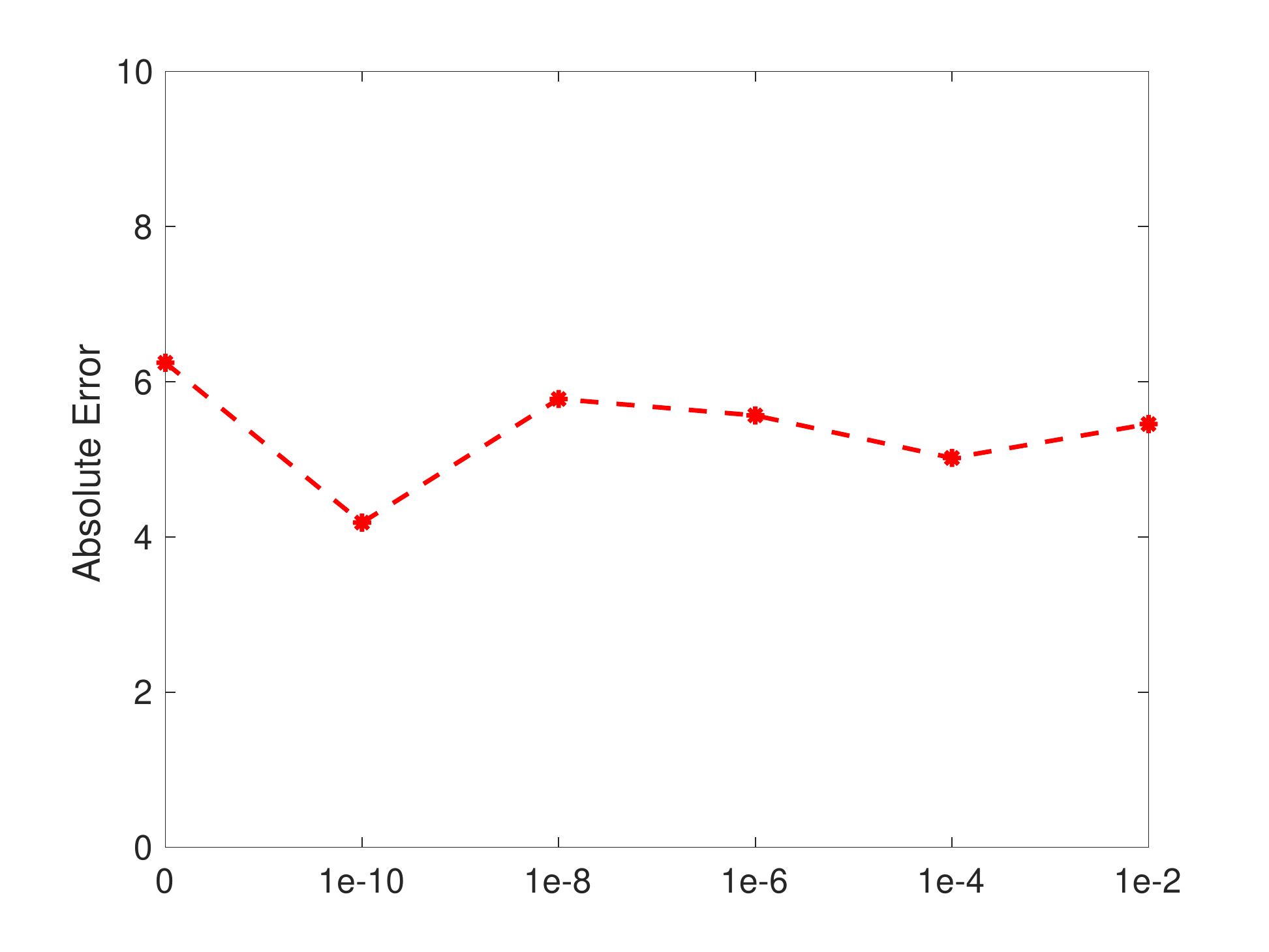}
  \caption{Results for News data.}
\end{subfigure}
\caption{Estimation error when varying $\gamma$.}
 \label{lr}
\end{figure}

To answer \textbf{RQ2}, we examine the impact of the imbalance penalty term in LTEE on the model performance.
In particular, we set the hyperparameter $\gamma \in \{0,10^{-10},10^{-8},10^{-6},10^{-4},10^{-2}\}$ and investigate how the estimation accuracy of LTEE varies. 
Results can be seen in Fig. \ref{lr}.
We begin by noting that indeed imbalance confers an advantage to using the W-1 regularization term, see e.g. the results for $\gamma=0$ and for $\gamma\neq 0$ using both datasets. We can also observe that when $\gamma$ gets too large, i.e., the imbalance penalty is overemphasized, LTEE's performance tends to degrade. The best performance is achieved at $\gamma=10^{-8}$ and $\gamma=10^{-10}$ for the IHDP and the News datasets, respectively. In general, LTEE is robust to $\gamma$, thus can be tuned for various applications in practice.
\subsubsection{Impact of the double heads}
\begin{figure}
\centering
\begin{subfigure}{.5\columnwidth}
\centering
  \includegraphics[width=\linewidth]{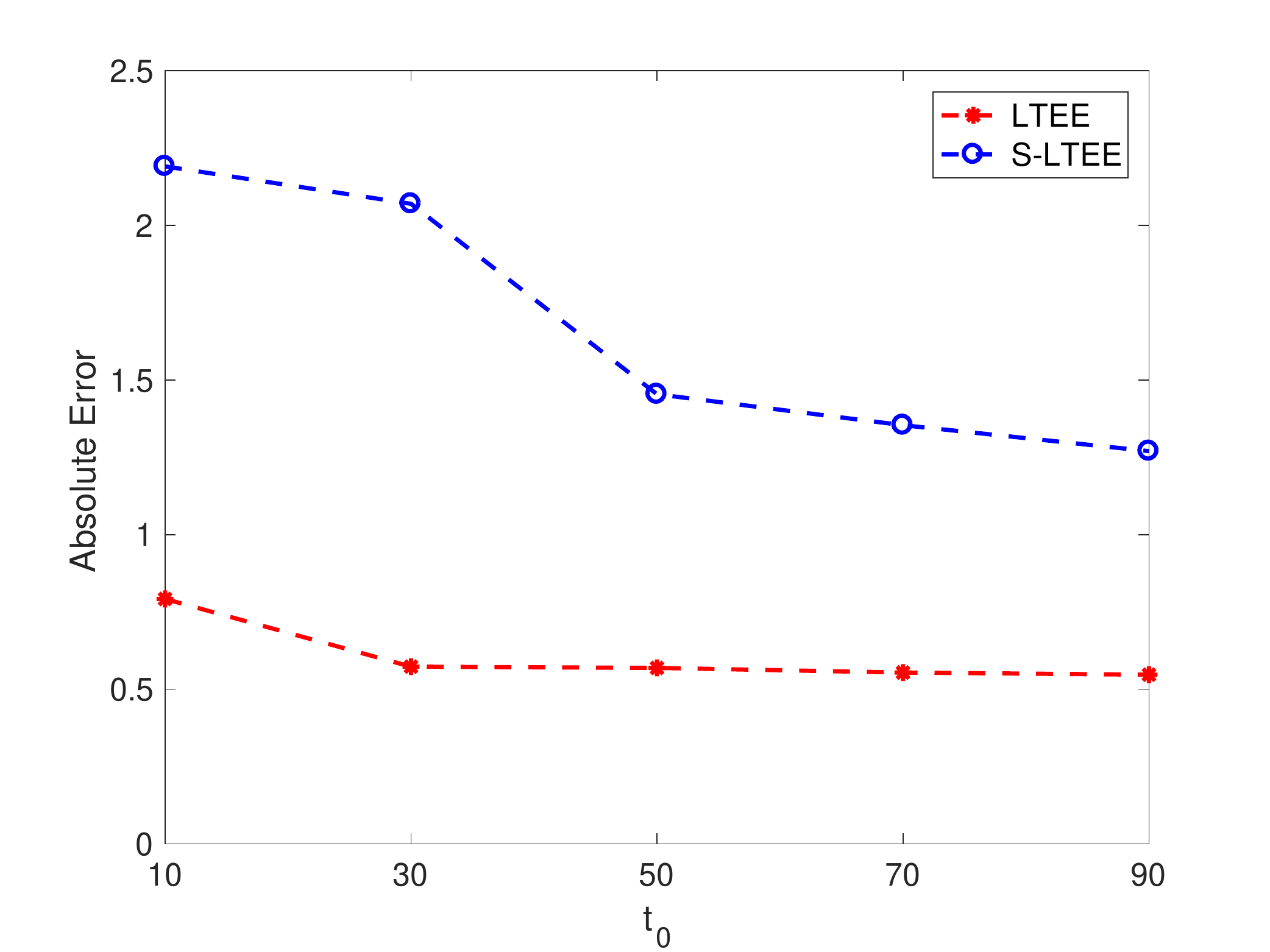}
  \caption{Varying $t_0$ and fixing $T$.}
\end{subfigure}%
\begin{subfigure}{.5\columnwidth}
\centering
  \includegraphics[width=\linewidth]{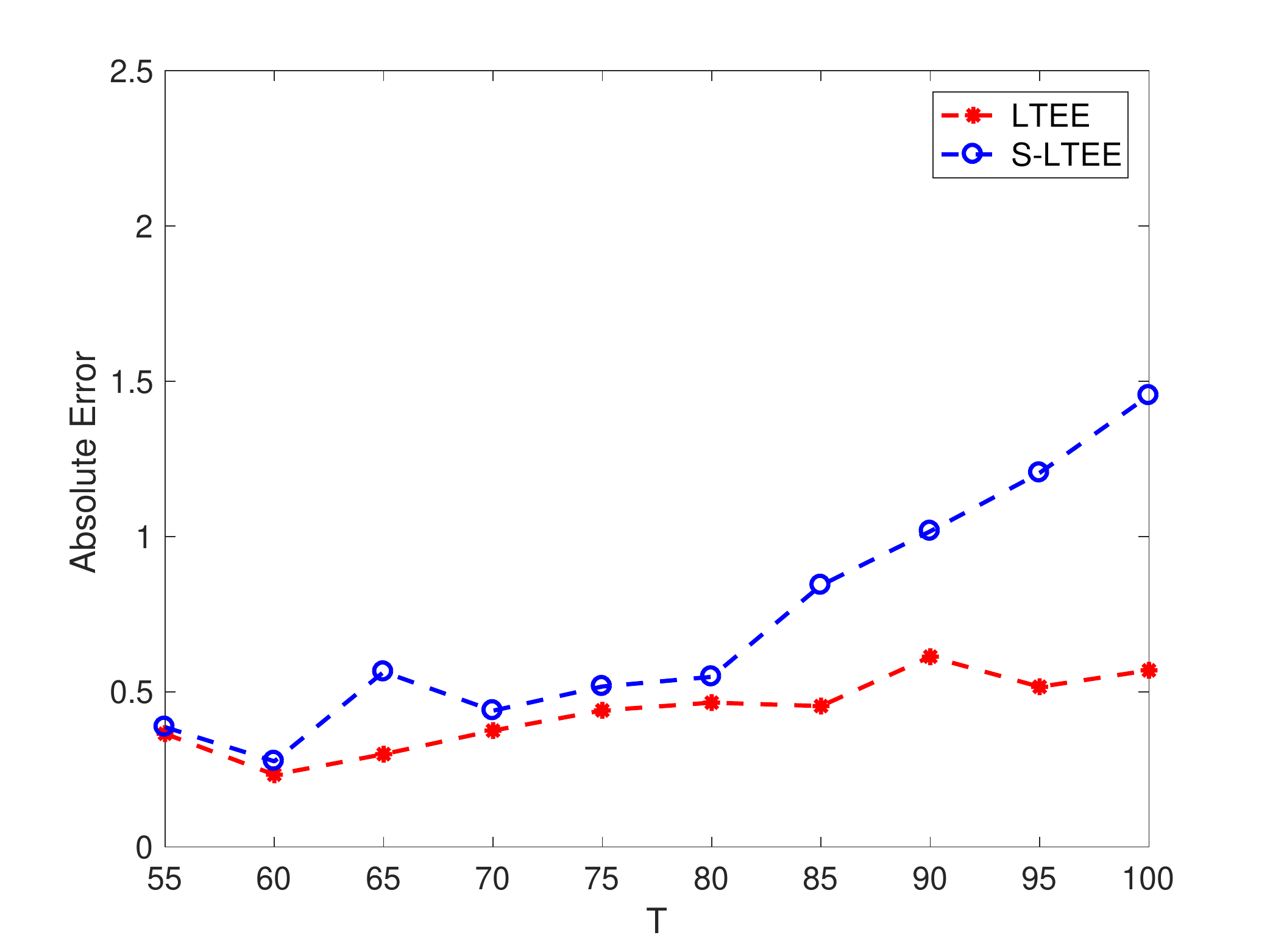}
  \caption{Varying $T$ and fixing $t_0$.}
\end{subfigure}
\caption{Comparisons of LTEE and  S-LTEE.}
 \label{single}
\end{figure}

To answer \textbf{RQ3}, we compare LTEE with its variant, the ``single-headed'' LTEE (S-LTEE) where model parameters are shared between the treated and control groups. Specifically, we estimate long-term effects under the aforementioned two scenarios (Sec. 4.2-4.3). In this (and next) ablation study, we report results for IHDP dataset as similar trends can be found using News dataset. We observe from Fig. \ref{single} that LTEE consistently achieves better performance than S-LTEE under both scenarios. We conclude that the double-headed architecture design contributes to improving the long term effects estimation in LTEE.
\subsubsection{Impact of the Nonlinearity}
\begin{figure}
\centering
\begin{subfigure}{.5\columnwidth}
\centering
  \includegraphics[width=\linewidth]{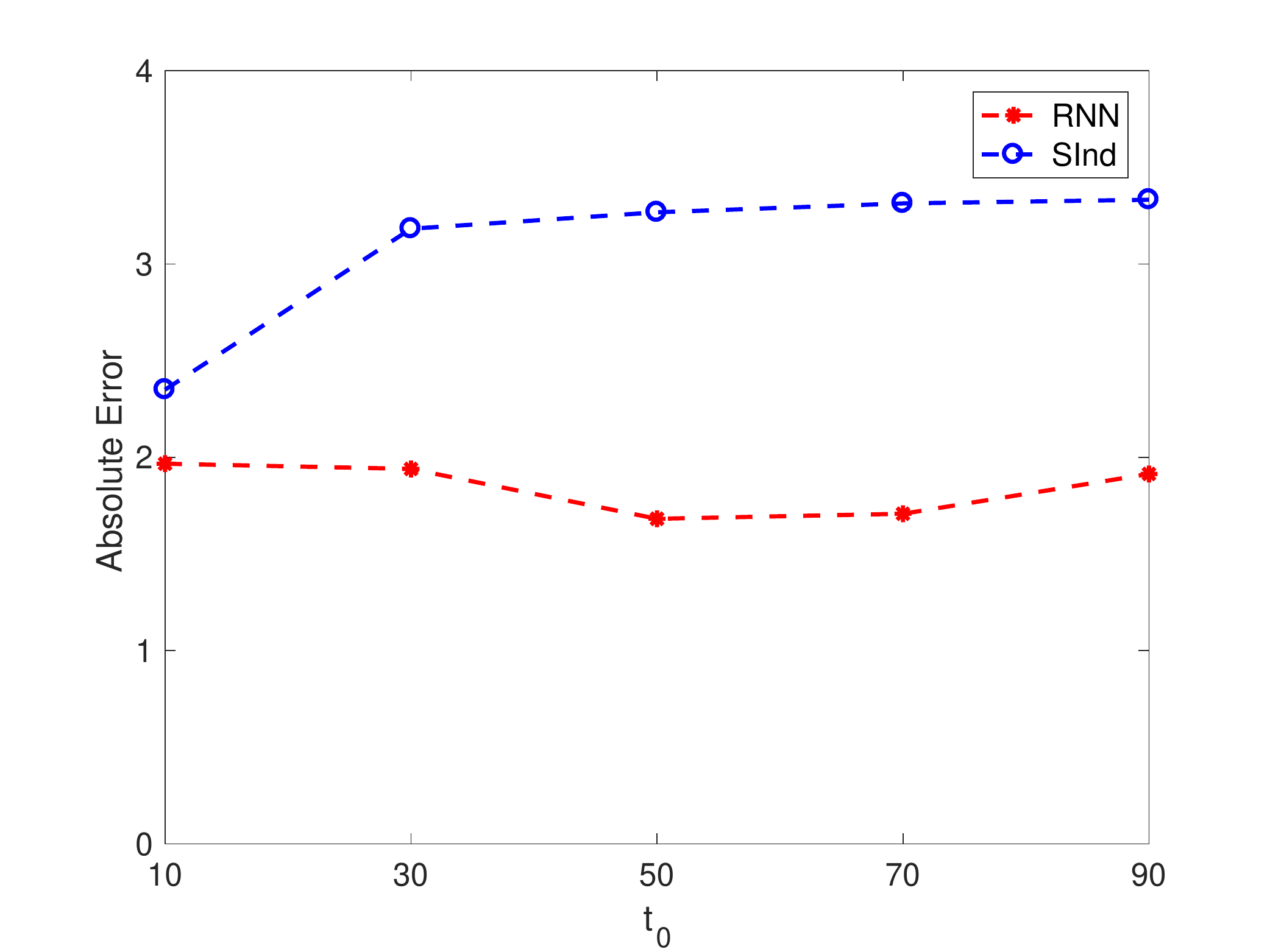}
  \caption{Varying $t_0$ and fixing $T$.}
\end{subfigure}%
\begin{subfigure}{.5\columnwidth}
\centering
  \includegraphics[width=\linewidth]{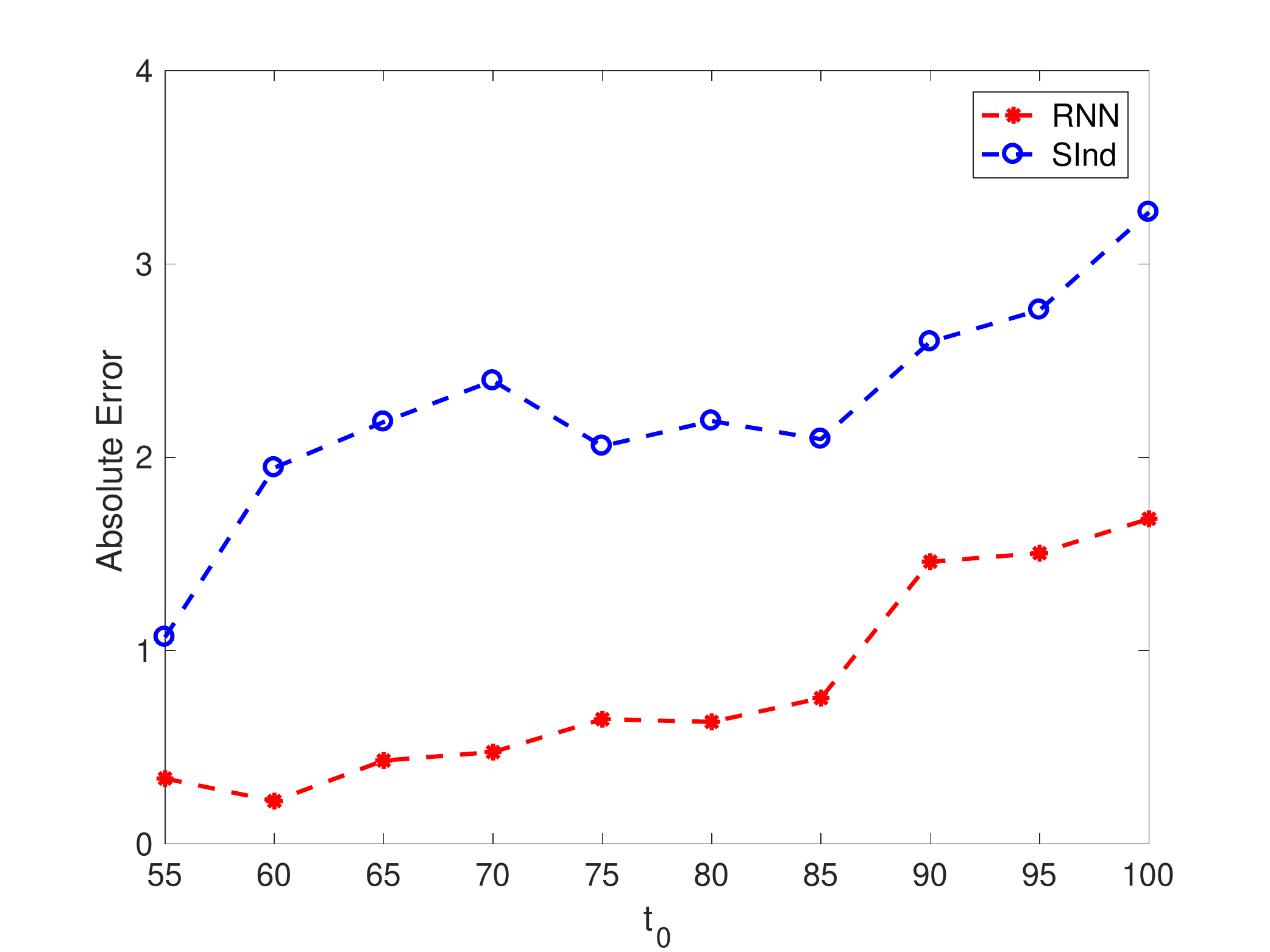}
  \caption{Varying $T$ and fixing $t_0$.}
\end{subfigure}
\caption{Comparisons of RNN and SInd.}
 \label{nonlinear}
\end{figure}

To answer \textbf{RQ. 4}, this experiment investigates whether nonlinearity alone in LTEE, i.e., the RNN, can outperform the state-of-the-art approach SInd, which assumes linear relationships among outcomes. We run RNN and SInd under scenarios in Sec. 4.2-4.3 using IHDP dataset. The consistent improvement of RNN over SInd as shown in Fig. \ref{nonlinear} demonstrates the superiority of the nonlinearity in LTEE. 

In summary, despite the fact that it does not use any short-term outcomes in the target data, LTEE outperforms established methods for estimating long-term causal effect when either the number of observed short-term outcomes ($t_0$) or the target endpoint ($T$) is fixed and the other varies. The improvement is mostly significant. These results evidence that the proposed surrogate representation can effectively address the major challenges in the two inferential tasks, including the stringent surrogacy assumption, the imbalance across treatment groups, the time invariant confounders and the overlook of time dependencies among observed outcomes.
\section{Related Literature}
Our work relates closely to missing data literature \cite{little2019statistical} and the literature in RNNs. In this section, we therefore survey along the following three dimensions: long-term effect estimation, missing value handling on time-series data, and RNNs.  
\subsection{Long-Term Effect Estimation}
To date, a common approach to estimating long-term causal effect is to analyze treatment effects on a short-term surrogate \cite{prentice1989surrogate}. However, this method relies on the stringent surrogacy assumption that is highly controversial in empirical applications. Under the surrogacy assumption, Chetty et. al \cite{chetty2011does} estimated long-term effect of class size and quality in grade kindergarten on children's lifelong earnings. They used children's earnings when they were in their mid-twenties as the estimates to predict effects on total lifelong earnings, assuming that the impacts on earnings in children's twenties would persist throughout their lives. A more recent work \cite{chetty2016effects} performed similar analysis on administrative data from tax returns to prove the impacts of Moving to Opportunity on children's primary outcomes (e.g., earnings).


A major concern of the surrogacy assumption is that the surrogate may not mediate the full effect of the treatment \cite{freedman1992statistical}. For example, when using test scores as the proxy of children's lifelong earnings, reductions in class size appear to affect earnings through changes in non-cognitive skills that cannot be fully captured by test scores \cite{chetty2011does}. In addition, the surrogacy assumption can be easily violated if any hidden/unmeasured confounders exist between the surrogate and primary outcome \cite{vanderweele2015explanation,frangakis2002principal}. There is some recent work seeking to address these issues \cite{athey2019surrogate,kallus2020role}. Athey et al. \cite{athey2019surrogate} proposed to use a set of intermediate outcomes, i.e., the \textit{surrogate index}, instead of one, to infer the primary outcome. While showing promising results, the ``hope'' that one of the surrogates in the surrogate index can satisfy the surrogacy assumption is still skeptical. Their model also requires short-term outcomes in the target data, which might not be available in reality. Kallus and Mao \cite{kallus2020role} relaxed the strong surrogate assumption by taking a missing-data approach under the assumption that the primary outcome is \textit{missing at random}. Another work \cite{toulis2016long} studied the long-term effect of policy changes in multiagent economies. The authors specifically considered the dynamic nature of agent's response to policy changes and aimed to estimate causal effects after an adaptation period. Central to this work is the behavioral game theory used to formulate the ignorability assumptions in causal inference. 
\subsection{Missing Value on Time-Series Data}
Time series data is common in nearly every domain, and issues with missing values often occur. In statistics, the process of replacing missing values with reasonable values is termed as imputation. An effective algorithm should make use of the time-series characteristics, i.e., the time dependencies. There are three missing data mechanisms: missing at random, missing completely at random and missing not at random. 
Common statistical approaches to the analysis of time-series data where some follow-up observations are missing are the Last Observation Carried Forward (LOCF) and Next Observation Carried Backward (NOCB). In LOCF, a missing follow-up visit value is replaced by a unit's previously observed value. Similarly, NOCB replaces a missing value with an observed value afterwards. Recent research shows that these two methods give biased estimate of the treatment effect and underestimates the variability of the estimated result \cite{molnar2008does,salim2008comparison}. The second method is interpolation, including deterministic and stochastic methods. One of the most common deterministic method is Nearest-Neighbor Interpolation \cite{sibson1981brief}. It is based on an intuitive idea: value of the closest known neighbor is assigned to replace the missing value. Another basic deterministic method is linear interpolation \cite{gnauck2004interpolation}, which searches for a straight line that passes through the end points. In stochastic methods, regression is commonly used for data imputation. While these methods are typically simple, studies show that they are tedious, inefficient and inapplicable \cite{chen2002interpolation,tripathi2007selection}. The third method is the combination of seasonal adjustment and linear interpolation, working well for data with both trend and seasonality \cite{brubacher1976interpolating}.
\subsection{Recurrent Neural Networks}
RNNs have been widely adopted in tasks with sequential data, such as text. The typical feature of RNN is a cyclic connection, enabling RNN to update the current state based on the past states and current input data. Traditional RNNs include fully RNN \cite{chen1996comparative}
and selective RNN \cite{vster2013selective}. The major issue with these RNNs is that they cannot connect the relevant information when the gap between input data is large. LSTM \cite{hochreiter1997long} is then proposed to handle the ``long-term dependencies'', and becomes the focus of deep learning architecture ever since. Due to the superior learning capacity, LSTMs have been widely used in various tasks including speech recognition \cite{hsu2016exploiting}, acoustic modeling \cite{qu2017syllable}, trajectory prediction \cite{altche2017lstm}, sentence embedding \cite{palangi2016deep}, correlation analysis \cite{mallinar2018deep} and time series forecasting \cite{petnehazi2019recurrent}. Recent research has witnessed a surge in applying RNNs as substitutes to many other machine learning and statistical techniques for time series forecasting. Most notably, a RNN achieved impressive performance in the recent M4 competition\footnote{https://forecasters.org/resources/time-series-data/m4-competition/}. Other successful architectures include GRU-RNN \cite{bahdanau2014neural}, DeepAR \cite{salinas2019deepar}, Multi-Quantile Recurrent Neural Network (MQRNN) \cite{wen2017multi}, Spline Quantile Function RNNs \cite{gasthaus2019probabilistic} and Deep State Space Models for probabilistic forecasting \cite{rangapuram2018deep}. While RNNs have shown promising results in time-series forecasting, it remains unclear how competitive RNNs can be in practice in an automated standard approach due to its complexity, interpretability and reproducibility \cite{petnehazi2019recurrent}. 
\section{Conclusion and Future Work}
This work studies the problem where short-term effect cannot be generalized to long-term effect, e.g., short-term ad clicks \textit{VS} long-term ad clicks. Long-term effect estimation consists of two inferential tasks that need to be performed simultaneously. Both tasks present their unique challenges. We therefore propose a principled framework, LTEE, that connects the concept of surrogate with sequential models in machine learning to learn \textit{surrogate representations} that (1) circumvent the strong surrogacy assumption; (2) redress the imbalance between different treatment groups; (3) allow time-varying confounders and (4) capture the time dependencies among observed outcomes. Extensive empirical evaluations corroborate the efficacy of the proposed approach.

Our work opens several key avenues for future research. First, the comparability assumption in LTEE needs to be further relaxed. There are real-world situations where the distribution of source data is not identical to that of target data. Second, limited access to real-world data restrains LTEE to be evaluated on semi-simulated data only. Future work can be directed towards deploying LTEE to online platforms where outcomes can be measured sequentially. Third, emerging literature \cite{toulis2016long} has defined the long-term causal effect as the effect after the system has stabilized. The adaptation of LTEE to estimating the stable effect needs further investigation. Another potential line of research can work on data with observed rather than inferred time-varying covariates of units. For studies where individuals need to take different treatments at different timesteps, e.g., clinical research, we can further extend LTEE to long-term effect estimation with time-varying treatments.
\section*{Acknowledgements}
This material is based upon work supported by the National Science Foundation (NSF) Grant 1614576.
\bibliographystyle{ACM-Reference-Format}
\bibliography{sample-base}









\end{document}